\documentclass[twocolumn,showpacs,aps,prl,letter,amsmath,amssymb,superscriptaddress]{revtex4-1}

\usepackage{graphicx}
\usepackage{dcolumn}
\usepackage{bm}
\usepackage{hyperref,soul}

\usepackage{epstopdf}

\usepackage{color,bbm}
\newcommand{\beq}{\begin{equation}}
\newcommand{\eeq}{\end{equation}}
\newcommand{\bqa}{\begin{eqnarray}}
\newcommand{\eqa}{\end{eqnarray}}
\newcommand{\nn}{\nonumber}

\newcommand{\rt}[1]{\sqrt{#1}\,}
\newcommand{\erf}[1]{Eq.~(\ref{#1})}

\newcommand{\ket}[1]{ |{#1} \rangle}

\newcommand{\sq}[1]{\left[ {#1} \right]}
\newcommand{\cu}[1]{\left\{ {#1} \right\}}

\newcommand{\an}[1]{\left\langle{#1}\right\rangle}
\newcommand{\tr}[1]{{\rm Tr}\sq{ {#1} }}

\newcommand{\paper}{Letter}
\newcommand{\xfrac}[2]{{#1}/{#2}}

\newcommand{\mf}{\mathbf}
\newcommand{\blk}{\color{black}}

\definecolor{maroon}{rgb}{0.7,0,0}

\definecolor{ngreen}{rgb}{0.3,0.7,0.3}

\definecolor{golden}{rgb}{0.8,0.6,0.1}

\begin{document}

\title{Experimental validation of quantum steering ellipsoids and tests of volume monogamy relations}

\author{Chao Zhang}
\affiliation{Key Laboratory of Quantum Information, University of Science and Technology of China, CAS, Hefei, 230026, People's Republic of China}
\affiliation{CAS Center For Excellence in Quantum Information and Quantum Physics, University of Science and Technology of China, Hefei, 230026, People's Republic of China}

\author{Shuming Cheng}
\email{shuming.cheng@griffithuni.edu.au}
\affiliation{Centre for Quantum Computation and Communication Technology (Australian Research Council), Centre for Quantum Dynamics, Griffith University, Brisbane, QLD 4111, Australia}

\author{Li Li}
\affiliation{Centre for Quantum Computation and Communication Technology (Australian Research Council), Centre for Quantum Dynamics, Griffith University, Brisbane, QLD 4111, Australia}

\author{Qiu-Yue Liang}
\affiliation{Key Laboratory of Quantum Information, University of Science and Technology of China, CAS, Hefei, 230026, People's Republic of China}
\affiliation{CAS Center For Excellence in Quantum Information and Quantum Physics, University of Science and Technology of China, Hefei, 230026, People's Republic of China}

\author{Bi-Heng Liu}
\affiliation{Key Laboratory of Quantum Information, University of Science and Technology of China, CAS, Hefei, 230026, People's Republic of China}
\affiliation{CAS Center For Excellence in Quantum Information and Quantum Physics, University of Science and Technology of China, Hefei, 230026, People's Republic of China}

\author{Yun-Feng Huang}
\email{hyf@ustc.edu.cn}
\affiliation{Key Laboratory of Quantum Information, University of Science and Technology of China, CAS, Hefei, 230026, People's Republic of China}
\affiliation{CAS Center For Excellence in Quantum Information and Quantum Physics, University of Science and Technology of China, Hefei, 230026, People's Republic of China}

\author{Chuan-Feng Li}
\email{cfli@ustc.edu.cn}
\affiliation{Key Laboratory of Quantum Information, University of Science and Technology of China, CAS, Hefei, 230026, People's Republic of China}
\affiliation{CAS Center For Excellence in Quantum Information and Quantum Physics, University of Science and Technology of China, Hefei, 230026, People's Republic of China}

\author{Guang-Can Guo}
\affiliation{Key Laboratory of Quantum Information, University of Science and Technology of China, CAS, Hefei, 230026, People's Republic of China}
\affiliation{CAS Center For Excellence in Quantum Information and Quantum Physics, University of Science and Technology of China, Hefei, 230026, People's Republic of China}

\author{Michael J. W. Hall }
\email{michael.hall@griffith.edu.au}
\affiliation{Centre for Quantum Computation and Communication Technology (Australian Research Council), Centre for Quantum Dynamics, Griffith University, Brisbane, QLD 4111, Australia}

\author{Howard M. Wiseman }
\email{h.wiseman@griffith.edu.au}
\affiliation{Centre for Quantum Computation and Communication Technology (Australian Research Council), Centre for Quantum Dynamics, Griffith University, Brisbane, QLD 4111, Australia}

\author{Geoff J. Pryde}
\email{g.pryde@griffith.edu.au}
\affiliation{Centre for Quantum Computation and Communication Technology (Australian Research Council), Centre for Quantum Dynamics, Griffith University, Brisbane, QLD 4111, Australia}

\date{\today}

\begin{abstract}
The set of all qubit states that can be steered to by measurements on a correlated qubit is predicted to form an ellipsoid---called the quantum steering ellipsoid---in the Bloch ball. This ellipsoid provides a simple visual characterization of the initial two-qubit state, and various aspects of entanglement are reflected in its geometric properties. We experimentally verify these properties via measurements on many different polarization-entangled photonic qubit states. Moreover, for pure three-qubit states, the volumes of the two quantum steering ellipsoids generated by measurements on the first qubit are predicted to satisfy a tight monogamy relation, which is strictly stronger than the well-known monogamy of entanglement for concurrence. We experimentally verify these predictions, using polarization and path entanglement. We also show experimentally that this monogamy relation can be violated by a mixed entangled state, which nevertheless satisfies a weaker monogamy relation.

\end{abstract}


\maketitle

{\it Introduction.---} The concept of \textit{steering} a quantum system, by means of measurement on a second system, was defined by Schr\"odinger~\cite{E35}. He showed that---in modern language---if two observers, Alice and Bob say, share an entangled pure state, then Alice, by making suitable measurements, can steer Bob's system to any desired state in the support of his local state, with a nonzero probability~\cite{E36}. This generalized the result by Einstein, Podolsky, and Rosen (EPR) that the ``real state of affairs'' for Bob, as described by his reduced state, appears to depend on actions carried out remotely by Alice~\cite{EPR35}. This ``spooky action at a distance'' led EPR to suggest that quantum mechanics cannot give a complete description of reality. It is now well known, however, that any attempt to give a local realistic model of quantum correlations must fail in some cases, due to the violation of Bell inequalities by some entangled quantum systems~\cite{B64,CHSH69,BCPSW14}. Moreover, it is precisely this failure---reflecting the fundamental nature of quantum steering---that has ultimately led to nonclassical information protocols with guaranteed security, such as quantum key cryptography~\cite{LCT14} and randomness generation~\cite{Random10}.

\begin{figure*}[!t]
	\centering
	\includegraphics[width=1\textwidth]{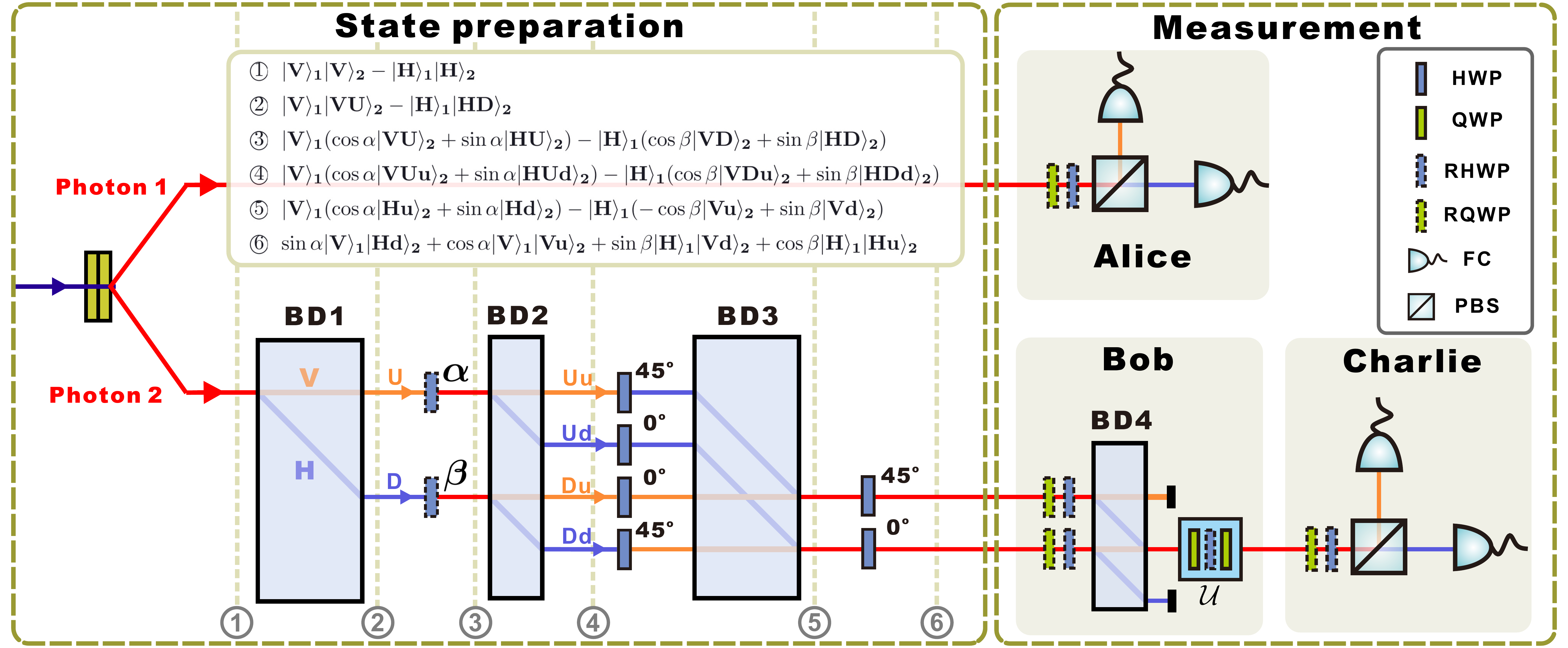}
	\caption{\label{Fig:1} Experimental setup. An ultraviolet laser pulse (centered at 390 nm) pumps a type-I SPDC source to generate a pair of polarization-entangled photons. Photon 1, a polarization qubit, is sent straight to Alice to perform projection measurements. Photon 2 passes through a series of beam-displacers (BD). The BDs cause the horizontally polarized ($H$) components (blue) to be walked off, while the vertically polarized ($V$) ones (orange) are transmitted undeviated, where BD1 and BD2 introduce a pair of path qubits, while BD3 eliminates the path qubit generated by BD1. We label the generated quantum state in each stage of state preparation setup  (normalization coefficients are omitted), see Supplemental Material~\cite{SM} for further details. Bob's qubit, the polarization of photon 2, is first measured by using BD4 and the rotatable wave plates. Meanwhile Charlie's qubit, encoded in the path degree of photon 2 (modes u and d), is transferred to the polarization degree of photon 2, which is then measured via the standard polarization analysis. The rotatable wave plates before BD4 are within a Mach-Zehnder interferometer, and they introduce a rotation-setting-dependent phase shift between its arms. Thus we implement another unitary operation ($\mathcal{U}$), to undo the effect of this phase shift at each setting. Each rotatable wave plate is mounted on a motorized rotation stage. Symbols used in the figure are HWP, half wave plate; QWP, quarter wave plate; RHWP, rotatable HWP; RQWP, rotatable QWP; FC, fiber-coupled detector; PBS, polarization beam splitter.}
\end{figure*}

Due to imperfections in physical state preparation and transmission, and protocols requiring the sharing of a quantum state between more than two parties, there is now substantial interest in the case that Alice and Bob do not share a pure state. In this more general scenario, entanglement is no longer sufficient for Alice to be able to steer Bob's system to any desired state, and a hierarchy of degrees of quantum correlation arises~\cite{GTM16}, starting with quantum discord at the bottom~\cite{OZ01,LV01} and rising through nonseparability~\cite{W89} and EPR steering~\cite{R89,WJD07,Saunders2010} to Bell nonlocality at the top~\cite{B64,CHSH69,BCPSW14}. Three important questions that arise in this scenario are these: which states can Alice can steer Bob's system to? What is the connection between this set of steered states and the degree of quantum correlation?  And for a multiparty state, are there restrictions on the degree to which one party can steer the systems of all other parties?

Surprisingly, only partial answers to the above questions are known, with most progress made for shared two-qubit~\cite{F02, MFCJ11,SSJYD12, JPJR14, ADST14,SMMMH15,NV16S} and three-qubit states~\cite{MJJWR14,CMHW16}. For a two-qubit state shared by Alice and Bob, it is theoretically predicted that \blk the set of Bob's steered states forms an ellipsoid in the Bloch  sphere~\cite{F02}. The geometric properties of this ellipsoid give necessary and sufficient conditions for the presence of discord and entanglement~\cite{JPJR14} and, for mixtures of Bell states, for EPR steerability~\cite{SMMMH15,NV16S}. For any pure three-qubit state shared by Alice, Bob, and Charlie, Bob's and Charlie's ellipsoids generated by Alice's measurements theoretically satisfy an elegant and tight volume monogamy relation~\cite{MJJWR14}.

In this \paper\ we first report the observation of the set of steered states for a variety of two-qubit states, and confirm their ellipsoidal nature by fitting experimental data to an ellipsoid equation via a least-square method with $R^2$ close to 1. Also, using photonic polarization and path entanglement, we are able to experimentally test the volume monogamy relation for three-qubit states. For very pure three-qubit states, we verify that the relevant volume monogamy relation is tight. Significantly, we also observe that, for a suitably prepared {\em mixed} entangled three-qubit state, this volume monogamy relation for pure states can be significantly violated, by 215 standard deviations.  However, a weaker monogamy relation~\cite{CMHW16}, valid for mixed states, is satisfied by all states in our experiment.

\begin{figure*}[!t]
	\centering
	\includegraphics[width=1\textwidth]{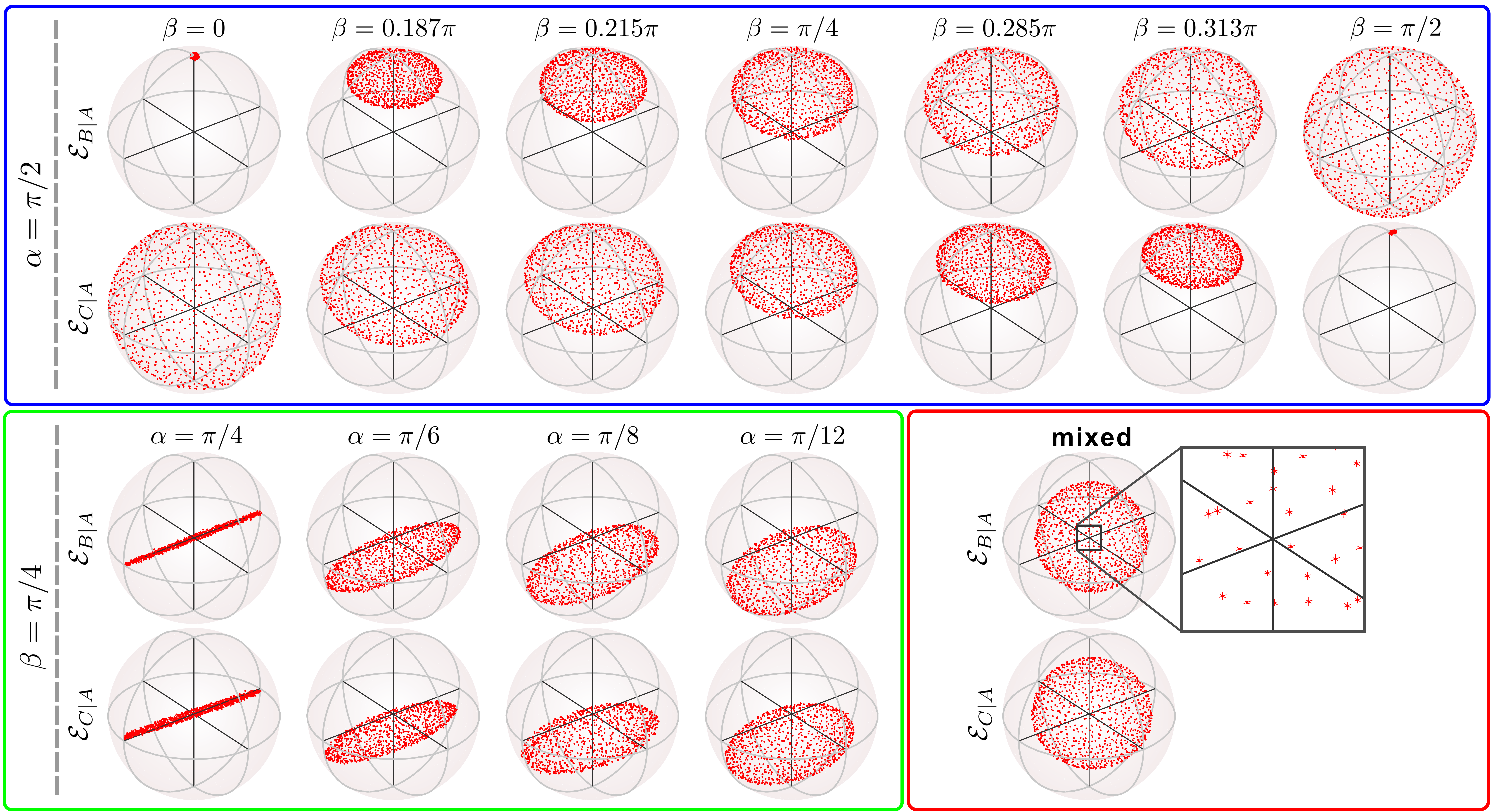}
	\caption{\label{Fig:2} The experimentally determined set of steered states (red points) for the family of pure three-qubit states~(Eq.~\ref{eq:family}) and the mixed state in Eq.~(\ref{eq:mix}). In the above blue box, $\alpha$ is fixed at $\pi/2$ and $\beta$ varies from $0$ to $\pi/2$---each entangled state belongs to the W class. In the green box, we fix $\beta=\pi/4$ and vary $\alpha$ from $\pi/4$ to $\pi/12$---the entangled states belong to the GHZ class. In the red box, $\rho_{ABC}$ in Eq.~(\ref{eq:mix}) is prepared. In each box, the upper figure refers to Bob's steered states while the lower one corresponds to Charlie's. For each state, we choose 1000 directions at random on the Bloch sphere~\cite{uniform} for Alice's measurements. Each red point, corresponding to Bob's or Charlie's steered state,  is reconstructed from $5.0\times10^4$ detection events via quantum state tomography. The bottom right inset shows the error bars (one line for each component) of measured red points, which has an average value of 0.007 for each component.}
\end{figure*}

{\it Quantum steering ellipsoids.---} A two-qubit state, shared by Alice and Bob, can be expressed in the standard Pauli basis $\boldsymbol{\sigma}\equiv (\sigma_1, \sigma_2, \sigma_3)$ as $ \rho_{AB}=1/4 \big(  \mathbbm{1}_A \otimes \mathbbm{1}_B+{\mf{a}} \cdot \boldsymbol{\sigma} \otimes \mathbbm{1}_B+\mathbbm{1}_A \otimes {\boldsymbol{b}} \cdot \boldsymbol{\sigma}  +\sum^3_{j,k=1}T_{jk}\sigma_j\otimes\sigma_k \big)$, where $\mathbbm{1}_A, \mathbbm{1}_B$ are identity operators. Here ${\mf{a}}$ and ${\mf{b}}$ are the Bloch vectors of Alice's and Bob's qubits, and $T$ is the spin correlation matrix. In terms of components ($j, k \in \cu{1, 2, 3}$), $a_j=\tr{\rho_{AB}\sigma_j\otimes \mathbbm{1}_B}, b_k=\tr{\rho_{AB}\mathbbm{1}_A\otimes \sigma_k}$ and $T_{jk}=\tr{\rho_{AB}\sigma_j\otimes \sigma_k}$.

When Alice makes a measurement on her qubit, each measurement outcome can be associated with an element $E\geq 0$ in a positive-operator valued measure (POVM) and thus assigned to an Hermitian operator $E=e_0\left(\mathbbm{1}_A+\mf{e}\cdot\boldsymbol{\sigma} \right)$ with $0\leq e_0\leq 1$ and $|\mf{e}|\leq 1$.  Correspondingly, Bob's qubit, correlated with Alice's, is steered to an unnormalized state $ {\rm Tr}_A[\rho_{AB}\, E\otimes \mathbbm{1}_B]$ with probability $p=\tr{\rho_{AB}\, E\otimes \mathbbm{1}_B}$. In particular, the normalized state admits the form $\frac{1}{2}\left[\mathbbm{1}_B+\xfrac{({\mf{b}}+T^\top\mf{e})\cdot \boldsymbol{\sigma}}{(1+{\mf{a}}\cdot\mf{e})}\right]$ for Bob's qubit. Then, considering all possible local measurements by Alice, this yields a set of Bob's steered states, represented by the set of Bloch vectors
\beq \label{reduced}
\mathcal{E}_{B|A}=\left\{ \frac{{\mf{b}}+T^\top\mf{e}}{1+{\mf{a}}\cdot\mf{e}}:|\mf{e}|\leq 1 \right\}.
\eeq
This set can be proven to form a (possibly degenerate) ellipsoid~\cite{F02}, and hence is called a quantum steering ellipsoid~\cite{JPJR14}. The subscript $B|A$ denotes Bob's steering ellipsoid generated by Alice's local measurements.

The quantum steering ellipsoid $\mathcal{E}_{B|A}$, together with the reduced Bloch vectors ${\mf{a}}$ and ${\mf{b}}$,  provides a faithful tool to visualize the two-qubit state~\cite{JPJR14}. The size of the steering ellipsoid can be quantified by its normalized volume $V_{B|A}=|\det (T-{\mf{a}}{\mf{b}}^\top)|/(1-|{\mf{a}}|^2)^2$~\cite{JPJR14}.
Here the volume is normalized relative to the total volume of the Bloch sphere, $\xfrac{4\pi}{3}$. It was found in Ref.~\cite{JPJR14} that the upper bound $V_{B|A}=1$ is achieved if and only if Alice and Bob share a pure entangled two-qubit state. In contrast, volumes of steering ellipsoids for all separable states are always no greater than $\xfrac{1}{27}$~\cite{JPJR14}.

{\it Volume monogamy relations.---} Consider the scenario where Alice, Bob, and Charlie share a  three-qubit state. The sets of steered states for Bob and Charlie, generated by measurements on Alice's qubit, could be described by the steering ellipsoids $\mathcal{E}_{B|A}$ and $\mathcal{E}_{C|A}$, respectively, which are further quantified by the volumes $V_{B|A}$ and $V_{C|A}$.

When the tripartite system is in a pure state $\ket{\psi_{ABC}}$, there exists a monogamy relation between volumes of steering ellipsoids~\cite{MJJWR14}
\beq
\rt{V_{B|A}}+\rt{V_{C|A}}\leq 1. \label{monogamypure}
\eeq
This relation is tight because it is nontrivially saturated if and only if $\ket{\psi_{ABC}}$ is a $W$-class state~\cite{MJJWR14,CMHW16}. Further, it is strictly stronger than the Coffman-Kundu-Wootters~(CKW) inequality for the concurrence measure of entanglement~\cite{CKW00}.

For mixed states, it is not possible to derive the above inequality, and below we experimentally produce a mixed state that violates~\erf{monogamypure}. However, some of us and a coworker derived a weaker monogamy relation~\cite{CMHW16}
\beq
({V_{B|A})^{\frac{2}{3}}+(V_{C|A})^{\frac{2}{3}}}\leq 1 \label{monogamymixed}
\eeq
which holds for all three-qubit states.  Both of these monogamy relations imply that Alice cannot steer both Bob and Charlie to a large set of states. For example, if Alice is able to steer Bob to the whole Bloch sphere (i.e., they share a pure entangled state), then Charlie's steering ellipsoid has zero volume (and indeed reduces to a single point).

\begin{figure}
\centering
\includegraphics[width=0.40\textwidth]{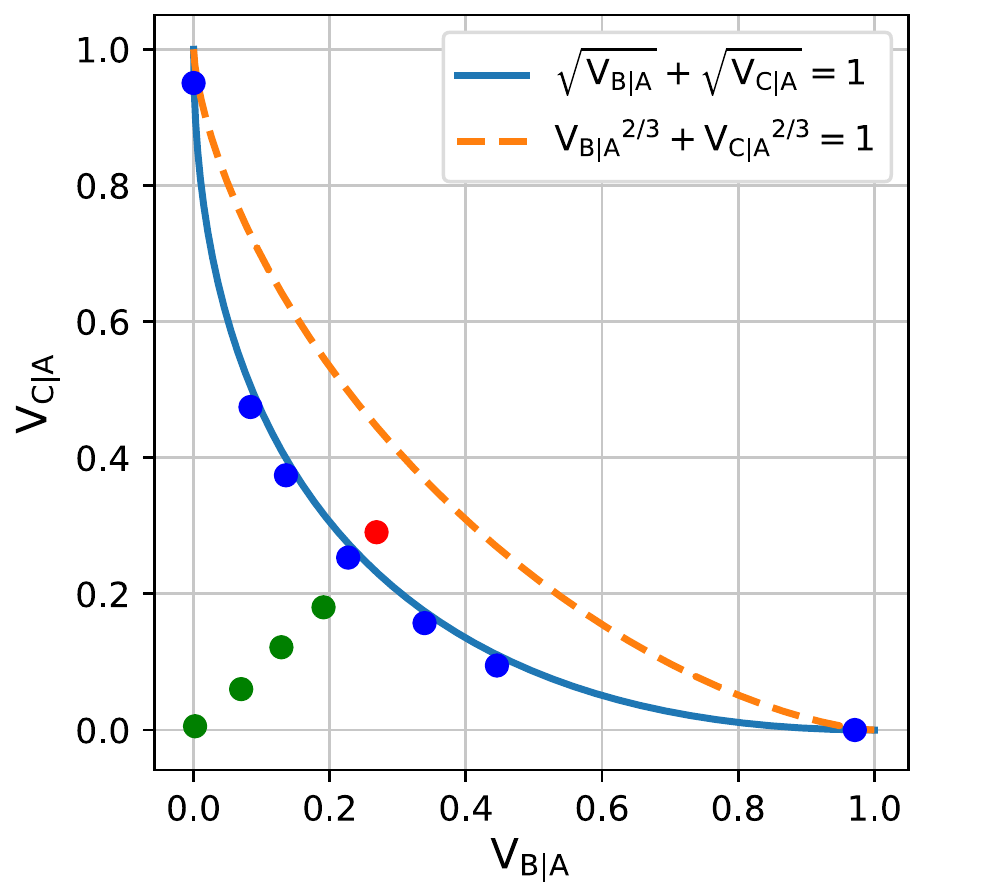}
\caption{\label{Fig:3} Volume monogamy relations. The x axis and y axis refer to the normalized volumes $V_{B|A}$ and $V_{C|A}$, respectively. The blue solid curve describes $\sqrt{ {V_{B|A}}}+\sqrt{  {V_{C|A}}}=1$, and the orange dashed curve represents $({V_{B|A}})^{2/3}+({V_{C|A}})^{2/3}=1$. Blue (green) points represent our experimental results for the $W$- (GHZ-) class states in the blue (green) box of Fig.~2. These blue points are almost located on the blue solid line, implying these states saturate the monogamy relation [Eq.~\eqref{monogamypure}] as predicted. The red point characterizes the measurement outcome for the mixed three-qubit state [Eq.~(\ref{eq:mix})]. It is sandwiched by the blue line and orange dashed line, indicating this state violates the monogamy relation [Eq.~(\ref{monogamypure})], but still satisfies the weaker one [Eq.~(\ref{monogamymixed})]. The error bars of the experimental data are of the order of $10^{-4}$, which is much smaller than the marker size.
}
\end{figure}

{\it Experimental setup.---} To experimentally verify the ellipsoidal nature [Eq.~(\ref{reduced})] and test the volume monogamy relation [Eq.~(\ref{monogamypure})], we first prepare a family of entangled three-qubit states which are well approximated by pure states of the form
\begin{eqnarray}
\label{eq:family}
|\psi_{ABC}\rangle &=& \frac{1}{\sqrt{2}}(\sin\alpha|100\rangle+\sin\beta|010\rangle \\\nonumber
& &+\cos\beta|001\rangle+\cos\alpha|111\rangle),
\end{eqnarray}
where $\alpha, \beta \in [0, \pi/2]$. In particular, when $\alpha=0, \pi/2$ or $\beta=0, \pi/2$, the state belongs to the set of $W$-class states; otherwise, the state belongs to the set of GHZ-class states~\cite{CMHW16, DVC01}. Further, this family is a good test bed for the monogamy relation as it covers the whole region enclosed by the inequality [Eq.~(\ref{monogamypure})], in addition to having a simple theoretical expression for the steering ellipsoids' volumes~\cite{CMHW16}.

The experimental setup to generate this family of states is shown in Fig.~1. First, we employ a type-I spontaneous parametric down-conversion (SPDC) source to produce a pair of polarization-entangled photons~\cite{KWWAE99}. Qubits $A$ and $B$ are encoded in the polarization of photon 1 and 2, respectively, while qubit $C$ is encoded in the path of photon 2. Then, a high-accuracy deterministic CNOT gate can be performed between qubit $B$ and qubit $C$ by using BDs and HWPs~\cite{L09,FAVRDW12}. Here we expand this to design and implement a sophisticated BD network that can produce the family of states [Eq.~(\ref{eq:family})] with tunable coefficients $\alpha,\beta$ as desired. The measurement process is shown in the right box of Fig.~1: Alice randomly chooses one direction on the Bloch sphere and performs the projection measurement on her qubit, while Bob and Charlie make measurements allowing single-qubit tomography of their individual qubits. After Alice has measured all sampled directions,  Bob's (Charlie's) steering ellipsoids can be verified by numerically fitting these tomographic data to an ellipsoid equation.

Within the same experimental setup, we also obtain the measurement statistics of a mixed entangled three-qubit state, which is predicted to violate the pure state monogamy relation in Eq.~\eqref{monogamypure}. This state is a mixture of two $W$-class states:
\begin{equation}
\label{eq:mix}
\rho_{ABC}=\frac{1}{2}(|\chi_1\rangle \langle \chi_1|+|\chi_2\rangle \langle \chi_2|),
\end{equation}
with
\begin{eqnarray}
\label{eq:mixstate}
|\chi_1\rangle=\frac{1}{\sqrt{6}}(|010\rangle-2|100\rangle+|001\rangle),\\\nonumber
|\chi_2\rangle=\frac{1}{\sqrt{6}}(|101\rangle-2|011\rangle+|110\rangle). \\\nonumber
\end{eqnarray}
State $\rho_{ABC}$ has purity$=\tfrac{1}{2}$. The state $|\chi_1\rangle$ is realized by first producing a nonmaximal entangled state $\sqrt{1/3}|00\rangle-\sqrt{2/3}|11\rangle$ in type-I SPDC and then setting $\alpha=\pi/2, \beta=\pi/4$ in the BD network~\cite{SM}. Noting that $|\chi_2\rangle$ could be generated from $|\chi_1\rangle$ by performing flipping operations $\sigma_x$ between states 0 and 1 for each qubit $|\chi_2\rangle=\sigma_x^{\otimes3}|\chi_1\rangle$, we only need to prepare the state $|\chi_1\rangle$ in the experiment, instead of $\rho_{ABC}$, because the measurement statistics of $\rho_{ABC}$ with respect to an arbitrary measurement $M$ are equal to those obtained by performing two measurements $M$ and $M'=\sigma_x^{\otimes3}M\sigma_x^{\otimes3}$ with equal probability on $|\chi_1\rangle$, see also Supplemental Material~\cite{SM}.

{\it Results.---} It is crucial in these experiments to prepare high fidelity tripartite entangled states. By employing states entangled in two degrees of freedom, we obtain the family of tripartite states [Eq.~(\ref{eq:family})] with nearly perfect fidelity and high generation rate. The state fidelity is calculated by $F=\langle \psi^{\rm  ideal}|\rho^{\rm  exp}|\psi^{\rm  ideal}\rangle$, where $\rho^{\rm  exp}$ is obtained via quantum state tomography. By carefully calibrating our setup, we achieve an average fidelity of 0.9887(1) for all of the prepared states (see Supplemental Material~\cite{SM} for more details) and the two-photon counting rate is about 6000 per second.

Our first result is to verify that the set of steered states for two-qubit systems indeed forms an ellipsoid. We use a nonlinear least-square method to fit our experimental data to an ellipsoid equation and employ $R^2$ to evaluate the fitting performances~\cite{GSN90}. The results are plotted in Fig.~2, and each steering ellipsoid is constructed via 1000 measurement points. As shown in Fig.~2, we have tested the steering shape for a variety of two-qubit states, generated from tracing out the qubit $B$ or qubit $C$ of three-qubit states. In particular, we observed that almost all $R^2$ parameters are close to unity~\cite{SM}, which confirms a good fit of our experimental points. For example, the smallest $R^2$ among all fitted ellipsoids (except degenerate cases) is 0.9956, corresponding to the ellipsoid $\mathcal{E}_{C|A}(\alpha=\pi/8, \beta=\pi/4)$. We also generated a family of pure two-qubit states with a varying degree of entanglement. We observe that the set of steered states closely coincides with the Bloch sphere for entangled two-qubit states and a single point on the surface for separable ones~\cite{SM}.

We next use the experimentally determined ellipsoids to test volume monogamy relations. Figure 3 plots the measured $V_{B|A}$ versus $V_{C|A}$ for all the 12 states of Fig.~\ref{Fig:2} \blk (see Supplemental Material~\cite{SM} for more details). In particular, for the $W$-class states in the blue box of Fig.~2, the corresponding $\sqrt{V_{B|A}}+\sqrt{V_{C|A}}$ range from $0.9754(3)$ to $0.9910(18)$, indicating that \blk these states nearly saturate the monogamy relation~(\ref{monogamypure}). For the GHZ-class states in the green box of Fig.~2, the measured pairs $(V_{B|A}, V_{C|A})$ all lie below the blue solid curve $\sqrt{V_{B|A}}+\sqrt{V_{C|A}}=1$. This can be used to classify different classes of three-qubit states by mapping the measured volume pair onto different regions of Fig.~3. It is interesting to point out that the steering ellipsoids for $W$-class states that saturate the volume monogamy relation also belong to a class of ``maximally obese'' states~\cite{JPJR14, ADST14,MJJWR14}, which have maximal volumes for the given centers.

Finally, it is  surprising to find that the suitably prepared mixed state [Eq.~(\ref{eq:mix})] violates the volume monogamy relation [Eq.~(\ref{monogamypure})] for pure states. The steering ellipsoids $\mathcal{E}_{B|A}$ and $\mathcal{E}_{C|A}$ are shown in the red box of Fig.~2, and the corresponding volume pair $(V_{B|A}, V_{C|A})$ is plotted as the red point in Fig.~3. The experimental values of $V_{B|A}$ and $V_{B|A}$ are $0.2688 (2)$ and $0.2906(2)$, respectively, which yields $\sqrt{V_{B|A}}+\sqrt{V_{C|A}}$ to be 1.0575(3). Nevertheless, this mixed state still satisfies a weaker monogamy relation given in Eq.~(\ref{monogamymixed}).

{\it Conclusions.---}We have experimentally verified the ellipsoidal nature of the set of steered states for a variety of two-qubit states (both two-photon states and one-photon states with two degrees of freedom). We used the experimentally determined ellipsoids to verify the monogamous nature of steering for a range of pure three-qubit states, and for mixed entanglement. It will be of both theoretical and experimental interest to investigate whether these distinct natures are still valid in more general scenarios. For example, can the volume monogamy relation for mixed three-qubit states be generalized to more than three parties? Is the ellipsoidal nature of the set of steered states valid for the qudit system beyond qubits? Can the monogamous nature of sets of steered states be confirmed in higher-dimensional multiparty systems? Finally, steering ellipsoids provide a powerful method to characterize quantum correlations of the system without shared reference frames, which may find further applications in the future quantum networks~\cite{WollmannPRA2018}. In the Supplemental Material~\cite{SM}, we investigate using just a small number of measurement settings to construct the steering ellipsoids.

{\it Acknowledgement.---}This work was supported by the National Key Research and Development Program of China (Grant No. 2017YFA0304100), the National Natural Science Foundation of China (Grants No. 61327901, No. 11774335, No. 11734015, No. 11474268, No. 11704371, and No. 11821404), Key Research Program of Frontier Sciences, CAS (Grant No. QYZDY-SSW-SLH003), the Fundamental Research Funds for the Central Universities (Grants No. WK2470000026 and No. WK2470000018), Anhui Initiative in Quantum Information Technologies (Grants No. AHY020100 and No. AHY070000), the National Youth Top Talent Support Program of National High-level Personnel of Special Support Program (Grant No. BB2470000005), China Postdoctoral Science Foundation (Grant No. 2017M612074), and the ARC Centre of Excellence (Grants No. CE110001027 and No. CE170100012). We thank Matt Palermo for useful discussions.

\appendix
	
\onecolumngrid
\setcounter{page}{1}
\renewcommand{\thepage}{Supplemental Material -- \arabic{page}/5}
\setcounter{equation}{0}
\renewcommand{\theequation}{S\arabic{equation}}

\section{SUPPLEMENTAL MATERIAL}

\subsection{Preparation of entangled 3-qubit states}

Here we show the detailed preparation of entangled 3-qubit states in Eq. (4) in the main text. First, a type-I SPDC source is employed to produce  a pair of polarisation-entangled photons, which can be expressed in the form \textcircled{1}
\begin{equation}
|\psi\rangle_{AB}=\sin\gamma |HH\rangle+\cos\gamma|VV\rangle, \label{polarisation}
\end{equation}
where $\gamma$ lies in the interval $[0, \pi]$. Here, H and V represent horizontal and vertical polarisations of two photons 1 and 2, respectively, while the subscripts A and B represent the polarisation qubits of the two photons.

Then, photon 2 is sent through an interferometer network which contains three beam displacers (BDs). In each BD, the H-polarised component experiences spatial walk-off, while the V-polarised component is transmitted undeflected. The thickness, and thus displacement distance, of BD1 and BD3 is double that of BD2 (the displacement distance of BD2 is 4 mm in our experiment). BD1 and BD2 will generate a pair of path qubits, labeled $L_1$ and $L_2$, on photon 2.  In the following (also in the main text), we label the upper (lower) paths of the photon introduced by BD1 and BD2 as U (D) and u (d) respectively. Thus, when photon 2 passes through BD1, the state~(\ref{polarisation}) becomes \textcircled{2}
\begin{equation}
|\psi\rangle_{ABL_1}=\sin\gamma |HHD\rangle+\cos\gamma|VVU\rangle.
\end{equation}
After BD1, the H-component passes through a HWP($\alpha$) and the V-component passes through a HWP($\beta$), which introduce the two tunable coefficients $\alpha, \beta$ in the state. And this process yields \textcircled{3}
\begin{equation}
|\psi\rangle_{ABL_1}=\cos\gamma |V\rangle(\sin\alpha|H\rangle+\cos\alpha|V\rangle)|U\rangle+\sin\gamma|H\rangle(\sin\beta|H\rangle+\cos\beta|V\rangle)|D\rangle.
\end{equation}
The path-dependent polarisation rotation can be regarded as a controlled-rotation operation between the path and polarisation qubits. When the photon passes through BD2, a second path qubit is introduced and now the state is \textcircled{4}
\begin{equation}
|\psi\rangle_{ABL_2L_1}=\cos\gamma |V\rangle(\sin\alpha|Hd\rangle+\cos\alpha|Vu\rangle)|U\rangle+\sin\gamma|H\rangle(\sin\beta|Hd\rangle+\cos\beta|Vu\rangle)|D\rangle.
\end{equation}

Finally, the polarisation qubit is rotated independently by setting HWPs oriented along either $45^\circ$ or $0^\circ$.  Note that $0^\circ$ HWPs are used to compensate the optical path difference and will introduce a $\pi$ phase shift on the state as soon as the V-component passes through it. Thus, we have a state \textcircled{5}
\begin{equation}
|\psi\rangle_{ABL_2L_1}=\cos\gamma |V\rangle|H\rangle(\sin\alpha|d\rangle+\cos\alpha|u\rangle)|U\rangle+\sin\gamma|H\rangle|V\rangle(\sin\beta|d\rangle-\cos\beta|u\rangle)|D\rangle.
\end{equation}
BD3 is then used to eliminate path qubit $L_1$. Thus, BD1 and BD3 form an Mach-Zehnder interferometer. Passing through the BD3, photons in the upper path further encounter a $45^\circ$ HWP, while photons in the lower path pass through a $0^\circ$ HWP. The output state is \textcircled{6}
\begin{equation}
|\psi\rangle_{ABL_2}=\cos\gamma (\sin\alpha|VHd\rangle+\cos\alpha|VVu\rangle)-\sin\gamma(\sin\beta|HVd\rangle+\cos\beta|HHu\rangle). \label{general class}
\end{equation}
By encoding the qubit modes H/V, D/U, d/u as logical state 0/1 and setting $\gamma=-45^\circ$, we obtain an entangled state that coincides with the family of 3-qubit states in Eq. (4) in the main text.

\subsection{Measurement statistics of $|\chi_1\rangle$ are sufficient}

It follows from Eq.~(\ref{general class}) that $|\chi_1\rangle$ in Eq. (6) in the main text can be first prepared in the experimental setup if
\beq
\sin\gamma=-\rt{\frac{1}{3}},~~\alpha=\frac{\pi}{2},~~\beta=\frac{\pi}{4}.
\eeq
Then, note that $|\chi_2\rangle$ could be generated from $|\chi_1\rangle$ by performing a swapping operation $\sigma_x$, which flips states 0 and 1, for each qubit, i.e.,
\beq
|\chi_2\rangle=\sigma_x\otimes\sigma_x\otimes\sigma_x|\chi_1\rangle.
\eeq
Thus, instead of preparing $\rho_{ABC}$ in Eq. (5) which is a equal mixture of $|\chi_1\rangle$ and $|\chi_2\rangle$, we only need to generate the state $|\chi_1\rangle$ in this experiment, because the measurement statistics of $\rho_{ABC}$ with respect to an arbitrary measurement $M$ are equal to those obtained by performing two measurements with equal probability on $|\chi_1\rangle$, i.e.,
\begin{align}
\an{M\rho_{ABC}}&=\frac{1}{2}\left(\an{M|\chi_1\rangle\langle\chi_1|}+\an{M|\chi_2\rangle\langle\chi_2|}\right) \nn\\
&=\frac{1}{2}\left(\an{M|\chi_1\rangle\langle\chi_1|}+\an{M\,\sigma_x\otimes\sigma_x\otimes\sigma_x|\chi_1\rangle\langle\chi_1|\sigma_x\otimes\sigma_x\otimes\sigma_x}\right) \nn \\
&= \frac{1}{2}\left(\an{M|\chi_1\rangle\langle\chi_1|}+\an{\sigma_x\otimes\sigma_x\otimes\sigma_x\,M\,\sigma_x\otimes\sigma_x\otimes\sigma_x|\chi_1\rangle\langle\chi_1|}\right) \nn \\
&\equiv \frac{1}{2}\left(\an{M|\chi_1\rangle\langle\chi_1|}+\an{M^\prime|\chi_1\rangle\langle\chi_1|}\right).
\end{align}
For the $\sigma_z$ and $\sigma_y$ measurements, $M'$ corresponds to flipping the outcomes of $M$, since $\sigma_x\sigma_z\sigma_x=-\sigma_z$ and $\sigma_x\sigma_y\sigma_x=-\sigma_y$. For the $\sigma_x$ measurement, $M'$ is equal to $M$.

\subsection{Experimental data fitting via the nonlinear least-square method}

We employ a nonlinear least-square method to fit the experimental points to verify the ellipsoidal nature of the set of steered states. All measured points are obtained via quantum state tomography on Bob's/Charlie's steered state, which can be faithfully represented by the Bloch vector $(x, y, z)$. Since there are 1000 samples, denote each data as a tuple $X_i=(x_i,y_i,z_i), i=1,\cdots,1000$. Then, we choose an ellipsoid equation to fit our experimental data, i.e.,
\beq
Y=f(X),
\eeq
where the fitting function $f$ is determined by the general ellipsoid equation, and we let $Y=z^2$.

Then we use the coefficient of determination $R^2$~\cite{Rsq} to evaluate how well experimental data are fitted. Specifically, we choose the Y-data to investigate the performance, and use
\begin{equation}
R^2\equiv 1-\frac{SS_{\rm res}}{SS_{\rm tot}}.
\end{equation}
Here $SS_{\rm res}=\sum_i(Y_i-f_i)^2$ refers to the sum of squares of residuals and $SS_{\rm tot}=\sum_i(Y_i-\bar{Y}_i)^2$ is the variance of the Y-data where $Y_i$ is the measured Y-data and $f_i$ is the corresponding fitted result. It is obvious that $R^2 \in [0, 1]$. More importantly, the better the fit is to the experimental data, the closer to unity $R^2$ is.

\subsection{Verification of steering ellipsoids for pure 2-qubit states}

\begin{figure*}[!t]
	\centering
	\includegraphics[width=0.94\textwidth]{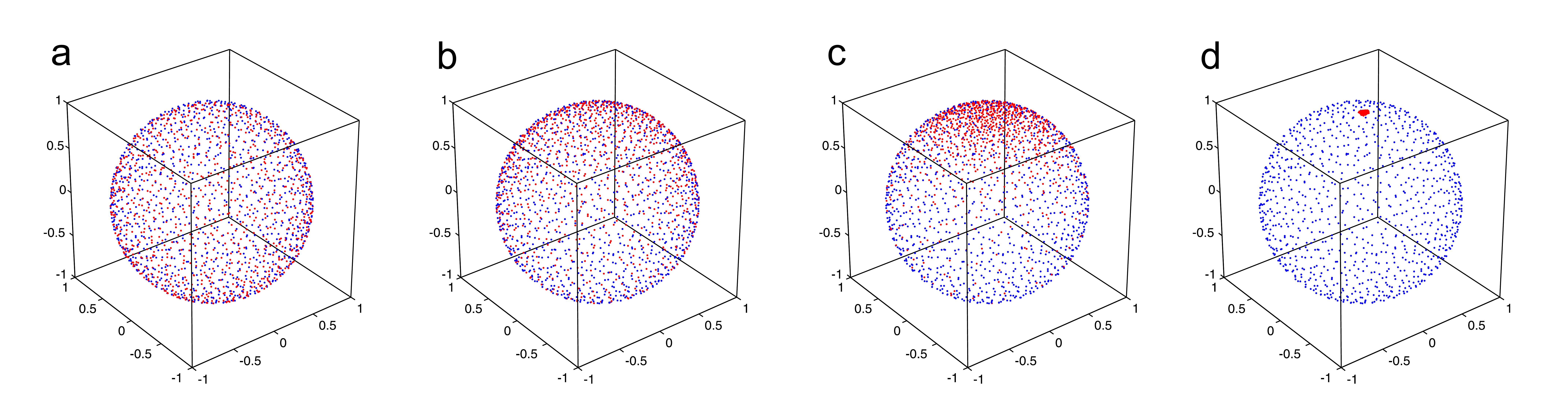}
	\caption{\label{Fig:S1}Quantum steering ellipsoids for a series of pure 2-qubit states, as per Eq.~(\ref{2qubit}), with $\gamma=\{\pi/4,\pi/6,\pi/12,0\}$. The red points are the experimentally-determined states for Bob, and the blue points represent the randomly-chosen measurement directions for Alice.}
\end{figure*}

In addition to the validation of quantum steering ellipsoids for the states generated from tracing out the qubit B or qubit C of 3-qubit states (Eq. (4)), we also prepare a class of pure 2-qubit states
\beq
|\psi\rangle_{AB}=\cos\gamma |HV\rangle+\sin\gamma|VH\rangle, \label{2qubit}
\eeq
using the type-I SPDC source directly. The measured steering ellipsoids are shown in Fig.~\ref{Fig:S1}. From left to right, the tested states correspond to $\gamma=\{\pi/4,\pi/6,\pi/12,0\}$ respectively. Our experimental results closely match the theoretical prediction that the set of steered states coincides with the Bloch sphere for entangled states, and a single point for separable states. Furthermore, it is worth noting that the uniform sampling of measurements for Bob may not lead to an uniformly distributed points on the steering ellipsoid, depending on the degree of entanglement.

\subsection{Test of the robustness of constructing ellipsoids with few measurement settings}

\begin{figure*}[!t]
	\centering
	\includegraphics[width=0.9\textwidth]{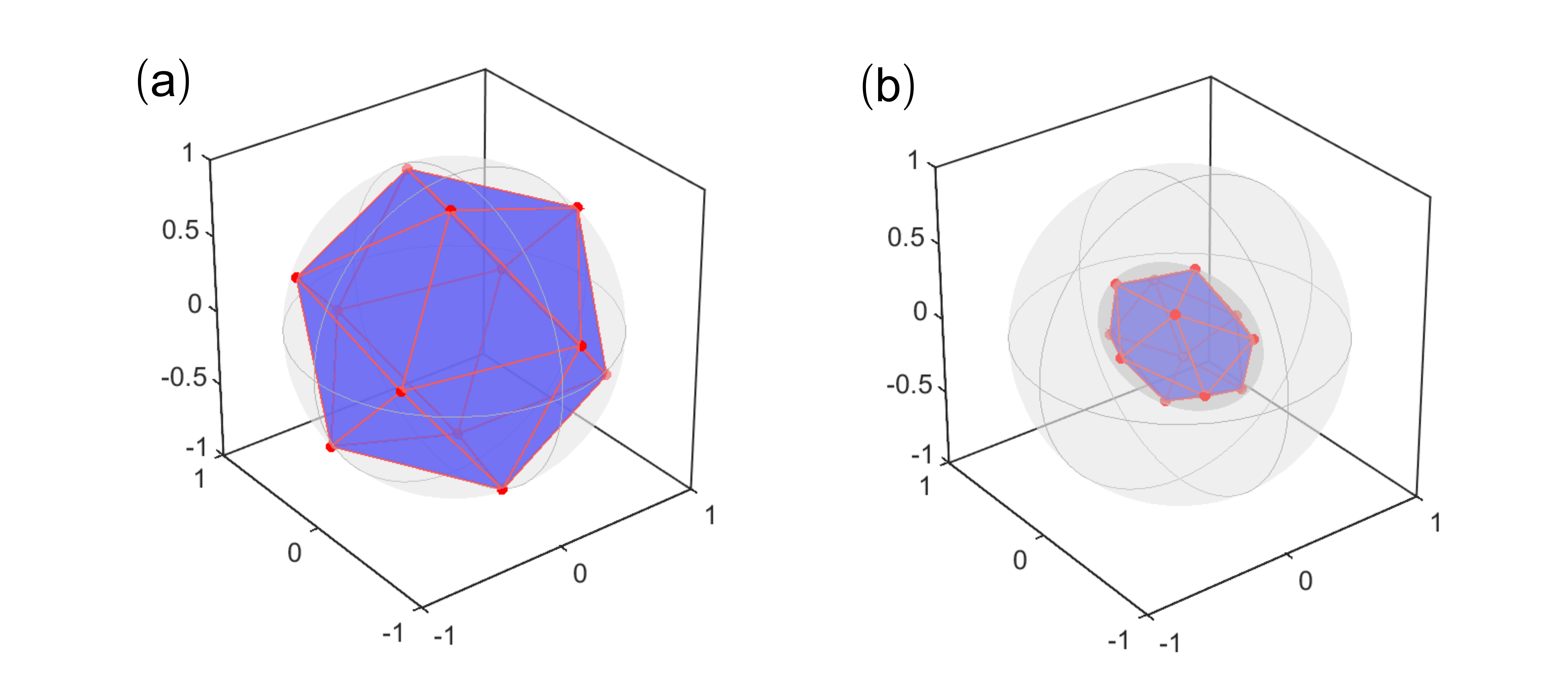}
	\caption{\label{Fig:S4}(a) The red points represent the measurement directions for Alice based on the icosahedron. (b) The red points represent the tomographic results for Bob's steered states.   }
\end{figure*}

In a potentially adversarial setting, the measurement directions should be selected randomly shot-by-shot. Thus, one always wants to use as few as possible measurement settings to construct the steering ellipsoid. It is known that a general ellipsoid is defined by a minimum of nine points. Here we consider measurement settings based around the platonic solids whose vertices are symmetric and uniformly distributed on the sphere. For example, the icosahedron has twelve vertices,  and this set is a good choice for Alice's measurement settings. To test the robustness of using only several points to construct the steering ellipsoid, we prepare a Bell-diagonal state which can be  written
\begin{equation}  \rho_{Bell}=0.6|\psi^-\rangle\langle\psi^-|+0.1|\psi^+\rangle\langle\psi^+|+0.1|\phi^-\rangle\langle\phi^-|+0.2|\phi^-\rangle\langle\phi^-|. \end{equation}
We first prepare the singlet state $|\psi^-\rangle$ state using the type-I source, and then apply a single-qubit gate (U) on one arm. This gate is chosen randomly from the set $\{I,\sigma_z,\sigma_x,\sigma_x\sigma_z\}$ with a probability distribution $\{0.6,0.1,0.1,0.2\}$, which leads to the Bell-diagonal state as desired when averaged over many runs. Then we perform 50 experiments. In each run of the experiment, we make a random rotation of the icosahedron,  and its vertices are used for Alice's measurements. Each instance of Bob's steered state is reconstructed from $5.0\times10^5$ detection events. Fig.~\ref{Fig:S4} shows the results of one experiment. We calculate the volume of the fitted ellipsoid for each of the 50 experiments, and obtain an average value of 0.0947 and a standard deviation of 0.0015. We also reconstruct steering ellipsoid by selecting only 9 of the 12 points in each experiment---the volume, $0.0946\pm 0.0016$, is consistent with the previous result. The small fluctuation for each run of the experiment (which is comparable with state tomography) demonstrate the validity of the ``icosahedron'' measurement strategy and alignment-free nature of the steering ellipsoids. The technique may find applications in future quantum networks to characterise quantum correlations without shared reference frames among distant parties.

\subsection{Data analysis}

\begin{figure*}[!t]
	\centering
	\includegraphics[width=1\textwidth]{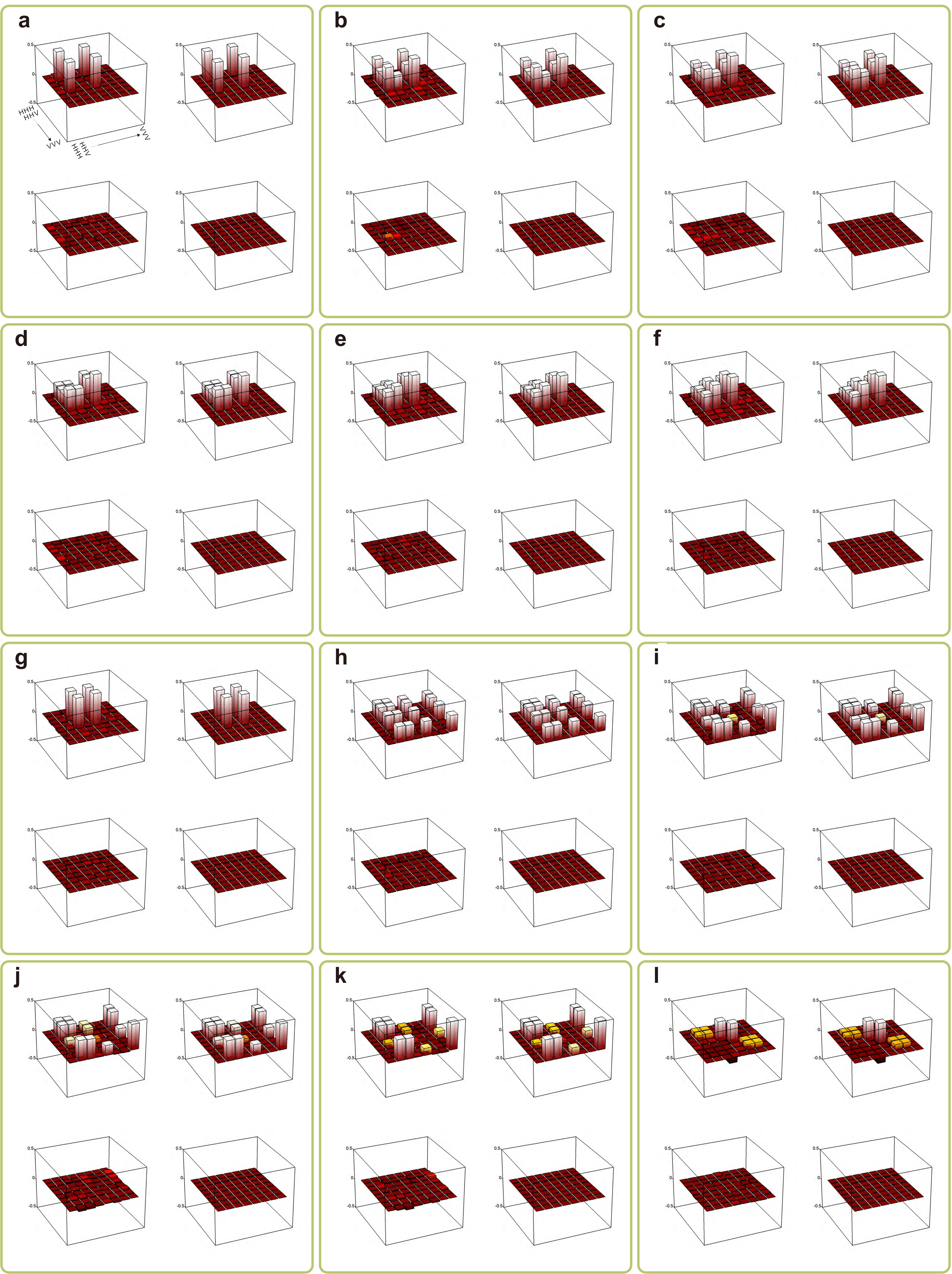}
	\caption{\label{Fig:S5}Tomographic results for all the tripartite states we have tested, labeled by a-l. In each box, the left two pictures show the real (top) and imaginary (bottom) part of the experimental reconstructed density matrix, while the right two pictures show the ideal density matrix.   }
\end{figure*}

Table.~\ref{Table:S1} shows the detailed fidelities and normalised volumes of the steering ellipsoids $\mathcal{E}_{B|A}$ and $\mathcal{E}_{C|A}$ for all the tripartite states we have tested. Fig.~\ref{Fig:S5} shows the tomographic results for each state.

There are two main imperfections in our system, one is due to the imperfection of the SPDC source and the other is the imperfection of the Mach-Zehnder interferences in the BD network. The noise is random and similar to the white noise, thus it has small contribution to the steering volumes. For the W-class states (a-g), the measured results of $\sqrt{V_{B|A}}+\sqrt{V_{C|A}}$ are always smaller than the theoretical value of 1. For the mixed state (l), the theoretical purity is 0.5, thus it is much different from the pure states we have tested.

\begin{table*}[!t!]
	\centering
	\caption{\label{Table:S1}The corresponding fidelities, the normalized volumes of steering ellipsoids and the goodness-of-fit parameters for all the tripartite states we have tested, labeled by a-l. The raw data have been corrected for the different detection efficiencies of the two fiber-coupled detectors of Alice and Charlie. The error bars are determined by Monte Carlo simulation (100 samples) with the photonic statistic error. exp, experimental; thy, theoretical; $SS_{\rm res}$, sum of squares of residuals; $R^2$, coefficient of determination. `Fidelity' means fidelity with the respective target state. }
	\begin{tabular}{|c|c|c|c|c|c|c|c|c|c|c|c|} \hline
		
		State & $\alpha$ & $\beta$ & Fidelity & $V_{B|A}^{\rm  exp}$ & $V_{C|A}^{\rm  exp}$ & $V_{B|A}^{\rm thy}$ & $V_{C|A}^{\rm thy}$ & $SS_{\rm res}(\mathcal{E}_{B|A})$ & $R^2(\mathcal{E}_{B|A})$ & $SS_{\rm res}(\mathcal{E}_{C|A})$ & $R^2(\mathcal{E}_{C|A})$\\\hline
		a&$\pi/2$ & 0             & 0.9914(2)    & 0.00004(1)   & 0.9504(5)    &0      &1         &0.00001&1.0   &0.1794&0.9978\\\hline
		b&$\pi/2$ & $0.187\pi$    & 0.9850(5)    & 0.0836(1)    & 0.4745(4)    &0.0944 &0.4800    &0.0091&0.9998 &0.0627&0.9992\\\hline
		c&$\pi/2$ & $0.215\pi$    & 0.9886(3)    & 0.1357(2)    & 0.3742(3)    &0.1528 &0.3710    &0.0141&0.9998 &0.0450&0.9995\\\hline
		d&$\pi/2$ & $\pi/4$       & 0.9905(3)    & 0.2271(3)    & 0.2533(3)    &0.25   &0.25      &0.0302&0.9996 &0.0283&0.9997\\\hline
		e&$\pi/2$ & $0.285\pi$    & 0.9910(3)    & 0.3393(3)    & 0.1571(2)    &0.3710 &0.1528    &0.0428&0.9995 &0.0165&0.9998\\\hline
		f&$\pi/2$ & $0.313\pi$    & 0.9885(3)    & 0.4456(4)    & 0.0948(1)    &0.4800 &0.0944    &0.0659&0.9991 &0.0091&0.9998\\\hline
		g&$\pi/2$ & $\pi/2$       & 0.9920(1)    & 0.9713(5)    & 0.00003(2)   &1      &0         &0.1065&0.9988 &0.00001&1.0\\\hline
		h&$\pi/4$ & $\pi/4$       & 0.9913(2)    & 0.0022(1)    & 0.0056(2)    &0      &0         &0.00003&0.3541 &0.0006&0.6304\\\hline
		i&$\pi/6$ & $\pi/4$       & 0.9890(3)    & 0.0699(2)    & 0.0601(2)    &0.0625 &0.0625    &0.0127&0.9979 &0.0130&0.9979\\\hline
		j&$\pi/8$ & $\pi/4$       & 0.9841(4)    & 0.1290(2)    & 0.1216(2)    &0.125  &0.125     &0.0223&0.9990 &0.0945&0.9956\\\hline
		k&$\pi/12$ & $\pi/4$      & 0.9847(4)    & 0.1909(3)    & 0.1803(3)    &0.1875 &0.1875    &0.0321&0.9993 &0.0385&0.9992\\\hline
		l&mixed &                 & 0.9888(3)    & 0.2688(2)    & 0.2906(2)    &0.2963 &0.2963    &0.0472&0.9968 &0.0417&0.9971\\\hline
		
	\end{tabular}
\end{table*}

\begin{thebibliography}{99}

\bibitem{E35}
E. Schr{\"o}dinger, Proc. Cambridge. Philos. Soc.  {\bf 31}, 555 (1935).

\bibitem{E36}
E. Schr{\"o}dinger, Proc. Cambridge. Philos. Soc. {\bf 32}, 446 (1936).

\bibitem{EPR35}
A. Einstein, B. Podolsky, and N. Rosen, Phys. Rev. {\bf 47}, 777 (1935).

\bibitem{B64} J. S. Bell, Physics  {\bf 1}, 195 (1964).

\bibitem{CHSH69} J. F. Clauser, M. A. Horne, A. Shimony, and R. A. Holt, Phys. Rev. Lett. {\bf 23}, 880 (1969).

\bibitem{BCPSW14} N. Brunner, D. Cavalcanti, S. Stefano, V. Scarani, and S. Wehner, Rev. Mod. Phys. {\bf 86}, 419 (2014).

\bibitem{LCT14} H. K. Lo, M. Curty, and K. Tamaki, Nat. Photonics {\bf 8}, 595 (2014).

\bibitem{Random10} S. Pironio, A.  Ac{\'{\i}}n, S. Massar, A. B. de La Giroday, D. N. Matsukevich, P. Maunz, S. Olmschenk, D. Hayes, L. Luo,  T. A. Manning, and C. Monroe, Nature (London) {\bf 464}, 1021 (2010).

\bibitem{GTM16} G. Adesso, T. R. Bromley, and M. Cianciaruso, J. Phys. A {\bf 49}, 473001 (2016).

\bibitem{OZ01} H. Ollivier and W. H. Zurek, Phys. Rev. Lett. {\bf 88}, 017901 (2001).

\bibitem{LV01} L. Henderson and V. Vedral, J. Phys. A {\bf 34}, 6899 (2001).

\bibitem{W89}  R. F. Werner, Phys. Rev. A {\bf 40}, 4277 (1989).

\bibitem{R89} M. D. Reid, Phys. Rev. A {\bf 40}, 913 (1989).
	
\bibitem{WJD07} H. M. Wiseman, S. J. Jones, and A. C. Doherty, Phys. Rev. Lett. {\bf 98}, 140402 (2007).

\bibitem{Saunders2010}  D. J. Saunders, S. J. Jones, H. M. Wiseman and G. J. Pryde,  Nat. Phys. \textbf{6}, 845 (2010)





\bibitem{F02} F. Verstraete, Ph.D. thesis, Katholieke Universiteit Leuven (2002).

\bibitem{MFCJ11} M. Shi, F. Jiang, C. Sun, and J. Du, New J. Phys. {\bf 13}, 073016 (2011).

\bibitem{SSJYD12} M. Shi, C. Sun, F. Jiang, X. Yan, and J. Du, Phys. Rev. A {\bf 85}, 064104 (2012).

\bibitem{JPJR14} S. Jevtic, M. Pusey, D. Jennings, and T. Rudolph, Phys. Rev. Lett. {\bf 113}, 020402 (2014).

\bibitem{ADST14} A. Milne, D. Jennings, S. Jevtic, and T. Rudolph, Phys. Rev. A {\bf 90}, 024302 (2014).

\bibitem{SMMMH15} S. Jevtic, M. J. W. Hall, M. R. Anderson, M. Zwierz, and H. M. Wiseman, J. Opt. Soc. Am. B {\bf 32}, A40 (2015).

\bibitem{NV16S} H. Chau Nguyen and T. Vu, Europhys. Lett. {\bf 115}, 10003 (2016).


\bibitem{MJJWR14} A. Milne, S. Jevtic, D. Jennings, H. Wiseman, and T. Rudolph, New J. Phys. {\bf 16}, 083017 (2014); ibid. {\bf 17}, 019501 (2015).

\bibitem{CMHW16} S. Cheng, A. Milne, M. J. W. Hall, and H. M. Wiseman, Phys. Rev. A {\bf 94}, 042105 (2016).

\bibitem{CKW00} V. Coffman, J. Kundu, and W. K. Wootters, Phys. Rev. A {\bf 61}, 052306 (2000).



\bibitem{DVC01} W. D\"ur, G. Vidal, and J. I. Cirac, Phys. Rev. A {\bf 62}, 062314 (2000).

\bibitem{SM} See Supplemental Material for details of state preparations, data fitting, and more results about steering ellipsoids, including the Ref.~\cite{Rsq}.

\bibitem{KWWAE99} P. G. Kwiat, E. Waks, A. G. White, I. Appelbaum, and P. H. Eberhard, Phys. Rev. A {\bf 60}, R773 (1999).

\bibitem{L09} B. P. Lanyon, M. Barbieri, M. P. Almeida, T. Jenewein, T. C. Ralph, K. J. Resch, G. J. Pryde, J. L. O'brien, A. Gilchrist, and A. G. White, Nat. Phys. {\bf 5}, 134 (2009).

\bibitem{FAVRDW12} O. J. Far\'{\i}as, G. H. Aguilar, A. Vald\'es-Hern\'andez, P. H. Souto Ribeiro, L. Davidovich, and S. P. Walborn, Phys. Rev. Lett. {\bf 109}, 150403 (2012).

\bibitem{GSN90} S. A. Glantz, B. K. Slinker, and T. B. Neilands, \emph{Primer of applied regression and analysis of variance} (McGraw-Hill, New York, 1990), Vol. 309.



\bibitem{uniform}J. Davies, Random points on a sphere, https://www.jasondavies.com/maps/random-points/.

\bibitem{Rsq} S. Wright, J. Agric. Res. {\bf 20}, 557 (1921).


\bibitem{WollmannPRA2018}
S. Wollmann, M. J. W. Hall, R. B. Patel, H. M. Wiseman and G. J. Pryde, Phys.\ Rev.\ A \textbf{98}, 022333 (2018).

\end{thebibliography}
\end{document}